\documentclass {cjpsuppl}         
\usepackage{graphicx}             

\begin{document}
\title{JORDAN-WIGNER APPROACH
       TO DYNAMIC CORRELATIONS IN 2D SPIN-$\frac{1}{2}$ MODELS\,\footnote{Presented
                                                   at 12-th Czech and Slovak Conference on Magnetism,
                                                   Ko\v{s}ice, 12-15 July 2004}}
\authori{Oleg Derzhko\,\footnote{Corresponding author. e-mail address: derzhko@icmp.lviv.ua}
         and Taras Krokhmalskii}
\addressi{Institute for Condensed Matter Physics NASU,\\
          1 Svientsitskii Street, L'viv-11, 79011, Ukraine}
\authorii{}     \addressii{}
\authoriii{}    \addressiii{}
\authoriv{}     \addressiv{}
\authorv{}      \addressv{}
\authorvi{}     \addressvi{}
%
\headauthor{Oleg Derzhko and Taras Krokhmalskii}
\headtitle{Jordan-Wigner approach \ldots}
\lastevenhead{Oleg Derzhko and Taras Krokhmalskii: Jordan-Wigner approach \ldots}
%

\refnum{0000 A}
\daterec{XXX}  
\suppl{A}  \year{2004}
\setcounter{page}{1}
\maketitle
\begin{abstract}
We discuss the dynamic properties
of the square-lattice spin-$\frac{1}{2}$ $XY$ model
obtained using the two-dimensional Jordan-Wigner fermionization approach.
We argue the relevancy of the fermionic picture
for interpreting the neutron scattering measurements
in the two-dimensional frustrated quantum magnet Cs$_2$CuCl$_4$.
\end{abstract}

Dynamic properties of two-dimensional quantum spin models
have attracted considerable attention
from both experimentalists and theoretists.
Among recent experimental studies in this field
the neutron scattering on Cs$_2$CuCl$_4$
\cite{[01],[02],[03],[04]}
seems to be very intriguing.
Cs$_2$CuCl$_4$ is a two-dimensional low-exchange quantum magnet.
The two-dimensional character arises owing to the layered crystal structure,
whereas the in-plane superexchange route mediated by two nonmagnetic Cl$^-$ ions
provides a small exchange interaction
($\sim 1- 4$ K)
between Cu$^{2+}$ ions
carrying a spin of $\frac{1}{2}$.
In each layer the exchange paths form a triangular lattice
with nonequivalent interactions
along chains ($J$) and along zig-zag bonds ($J^\prime$).
The minimal Hamiltonian which determines the magnetic order
is the two-dimensional Heisenberg Hamiltonian
\begin{eqnarray}
\label{01}
H=\sum_{\langle i,i^{\prime}\rangle}J{\bf{s}}_i\cdot{\bf{s}}_{i^{\prime}}
+\sum_{\langle i,j\rangle}J^{\prime}{\bf{s}}_i\cdot{\bf{s}}_j,
\end{eqnarray}
where the first term describes the noninteracting chains
($J=0.374(5)$ meV)
and the second one describes the interchain coupling
($J^{\prime}=0.34(3) J$).
The interlayer coupling is small
$J^{\prime\prime}=0.045(5) J$
and it stabilizes the long-range magnetic order
below $T_{{\rm{N}}}=0.62(1)$ K.
The neutron scattering measurements in the spin liquid phase
(i.e. above $T_{{\rm{N}}}$ but below $J$, $J^\prime$
when the two-dimensional magnetic layers are decoupled)
clearly indicate
that the dynamic correlations are dominated
by highly dispersive excitation continua
which is a characteristic signature of fractionalization of spin-$1$ spin waves
into pairs of deconfined spin-$\frac{1}{2}$ spinons.
Linear spin-wave theory including one- and two-magnon processes
cannot describe the continuum scattering.

Two theories of the dynamical correlations
in the spin liquid phase in Cs$_2$CuCl$_4$
have so far been proposed.
A quasi-one-dimensional approach \cite{[05]}
starts from antiferromagnetic Heisenberg chains
along the strongest-exchange direction in the two-dimensional triangular planes
(thus immediately introducing spinon language)
and perturbatively treats all other terms in the Hamiltonian.
Note, however, that the coupling between chains is relatively large.
Another approach is based on the explicitly two-dimensional resonating-valence-bond picture
in which spinons
appear to be two spin-$\frac{1}{2}$ ends of a broken singlet bond
that separate away through bond rearrangement \cite{[06],[07]}.

\begin{figure}[t]
\vspace{-10mm}
\begin{center}
\includegraphics[width=90mm]{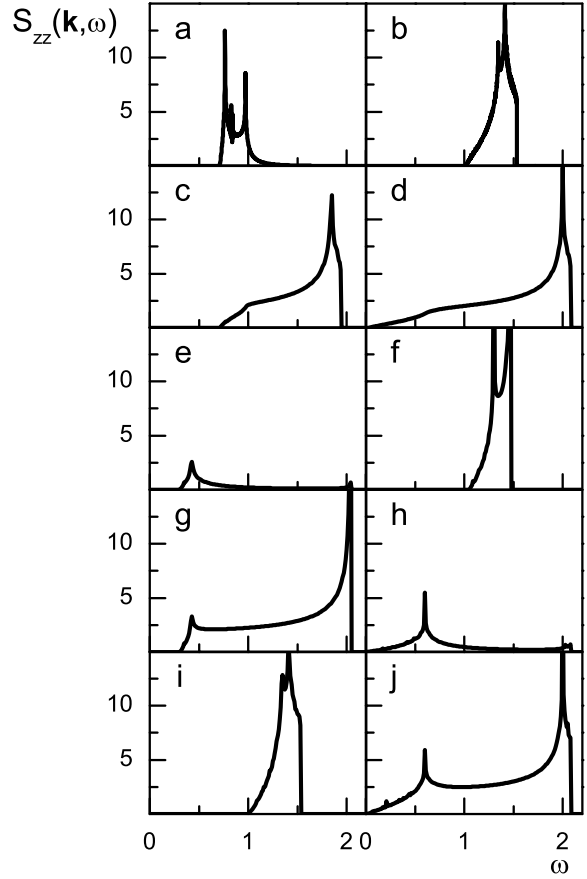}           
\end{center}
\vspace{-5mm}
\caption
{$S_{zz}({\bf{k}},\omega)$ (\ref{06})
of the square-lattice spin-$\frac{1}{2}$ $XY$ model (\ref{02})
with $J=1$, $J_\perp=0.3$ at zero temperature.
Frequency shapes are shown along
$k_x=\frac{\pi}{4}$, $k_y=0$ (a),
$k_x=\frac{\pi}{2}$, $k_y=0$ (b),
$k_x=\frac{3\pi}{4}$, $k_y=0$ (c),
$k_x=\pi$, $k_y=0$ (d),
$k_x=0$, $k_y=\frac{\pi}{2}$ (e),
$k_x=\frac{\pi}{2}$, $k_y=\frac{\pi}{2}$ (f),
$k_x=\pi$, $k_y=\frac{\pi}{2}$ (g),
$k_x=0$, $k_y=\pi$ (h),
$k_x=\frac{\pi}{2}$, $k_y=\pi$ (i),
$k_x=\pi$, $k_y=\pi$ (j).
\label{fig01}}
\end{figure}

In what follows
we discuss the two-dimensional Jordan-Wigner fermionization approach
to dynamic correlations \cite{[08]}
which eventually may capture the physics in spin liquid phase as well.
To be specific,
we consider a simpler square-lattice spin-$\frac{1}{2}$ $XY$ model
defined by the Hamiltonian
\begin{eqnarray}
\label{02}
H=\sum_{i=0}^\infty\sum_{j=0}^\infty
\left[
J\left(s_{i,j}^xs_{i+1,j}^x+s_{i,j}^ys_{i+1,j}^y\right)
+
J_{\perp}\left(s_{i,j}^xs_{i,j+1}^x+s_{i,j}^ys_{i,j+1}^y\right)
\right].
\end{eqnarray}
Here $J$ ($J_\perp$) is the exchange interaction in the horizontal (vertical) direction.
We are interested in the $zz$ dynamic structure factor
\begin{eqnarray}
\label{03}
S_{zz}({\bf{k}},\omega)
=
\sum_{p=0}^\infty\sum_{r=0}^\infty
\exp\left[{\mbox{i}}\left(k_xp+k_yr\right)\right]
\int_{-\infty}^\infty{\mbox{d}}t
\exp\left({\mbox{i}}\omega t\right)
\nonumber\\
\cdot
\left(
\langle s_{n,m}^z(t)s_{n+p,m+r}^z\rangle
-
\langle s_{n,m}^z\rangle\langle s_{n+p,m+r}^z\rangle
\right).
\end{eqnarray}
Performing the transformation to spinless fermions \cite{[09]},
\begin{eqnarray}
\label{04}
s_{i,j}^-=\exp\left({\mbox{i}}\alpha_{i,j}\right)d_{i,j},
\;\;\;
\alpha_{i,j}=\sum_{f=0(\ne i)}^\infty \sum_{g=0(\ne j)}^\infty
\Im \ln\left[f-i+{\mbox{i}}\left(g-j\right)\right]
d^\dagger_{f,g}  d_{f,g},
\end{eqnarray}
and adopting the mean-field treatment of the phase factors
which appear in (\ref{02})
we arrive at the Hamiltonian
\begin{eqnarray}
\label{05}
H=\frac{1}{2}\sum_{i=0}^\infty\sum_{j=0}^\infty
\left[
\left(-1\right)^{i+j}
J\left(d_{i,j}^\dagger d_{i+1,j}-d_{i,j}d_{i+1,j}^\dagger\right)
+
J_{\perp}\left(d_{i,j}^\dagger d_{i,j+1}-d_{i,j}d_{i,j+1}^\dagger\right)
\right],
\end{eqnarray}
which can be diagonalized by the Fourier and Bogolyubov transformations.
The $zz$ correlation function which enters (\ref{03})
is the density-density correlation function in fermionic picture (\ref{05});
it can be calculated with the help of the Wick-Bloch-de Dominicis theorem.
As a result we get
\begin{eqnarray}
\label{06}
S_{zz}({\bf{k}},\omega)
=
\int\frac{{\mbox{d}}{\bf{q}}}{4\pi}
\left\{
C^2
\left[
n_{{\bf{q}}}\left(1-n_{{\bf{q}}+{\bf{k}}}\right)
\delta\left(\omega+\Lambda_{{\bf{q}}}-\Lambda_{{\bf{q}}+{\bf{k}}}\right)
\right.
\right.
\nonumber\\
\left.
\left.
+\left(1-n_{{\bf{q}}}\right)n_{{\bf{q}}+{\bf{k}}}
\delta\left(\omega-\Lambda_{{\bf{q}}}+\Lambda_{{\bf{q}}+{\bf{k}}}\right)
\right]
+\left(1-C^2\right)
\left[
n_{{\bf{q}}}n_{{\bf{q}}+{\bf{k}}}
\delta\left(\omega+\Lambda_{{\bf{q}}}+\Lambda_{{\bf{q}}+{\bf{k}}}\right)
\right.
\right.
\nonumber\\
\left.
\left.
+
\left(1-n_{{\bf{q}}}\right)\left(1-n_{{\bf{q}}+{\bf{k}}}\right)
\delta\left(\omega-\Lambda_{{\bf{q}}}-\Lambda_{{\bf{q}}+{\bf{k}}}\right)
\right]
\right\},
\end{eqnarray}
where
$C=\cos\frac{\gamma_{{\bf{q}}+{\bf{k}}}-\gamma_{{\bf{q}}}}{2}$,
$\cos \gamma_{\bf{k}}=\frac{J_\perp \cos k_y}{\Lambda_{\bf{k}}}$,
$\Lambda_{\bf{k}}=\sqrt{J^2\sin^2k_x+J_\perp^2\cos^2k_y}$
and $n_{\bf{k}}$ is the Fermi function.
The $zz$ dynamic structure factor (\ref{06})
similarly to the one-dimensional cases \cite{[10]}
is conditioned by particle-hole pair excitations.
This two-fermion excitation continuum
implies nonzero values of $S_{zz}({\bf{k}},\omega)$
only in the restricted ${\bf{k}}$--$\omega$ region
with a sharp high-frequency cutoff.
Within the excitation continuum
$S_{zz}({\bf{k}},\omega)$ exhibits several washed-out excitation branches,
in particular,
the spin wave \cite{[08]}.
The representative scans throughout the Brillouin zone
of the zero-temperature $S_{zz}({\bf{k}},\omega)$
plotted in Fig. \ref{fig01}
show an extended structure
and, at least at a qualitative level,
strongly resemble some features of the frequency shapes
measured by neutron scattering in Cs$_2$CuCl$_4$
(see
Fig. 6 of Ref. \cite{[04]}).

To summarize,
we have computed the $zz$ dynamic structure factor
of the square-lattice spin-$\frac{1}{2}$ $XY$ model
using the two-dimensional Jordan-Wigner fermionization approach.
The dynamic structure factor exhibits both
sharp particle-like peaks and extended continuum profiles
in the constant wave vector
frequency shapes
which arise owing to particle-hole excitations.
Such features which have been observed experimentally in Cs$_2$CuCl$_4$
appear immediately within the two-dimensional Jordan-Wigner fermionization approach
that is obviously an attractive feature of this scheme.
One may also expect that in the limit of small interchain interaction this approach works well
since it contains the exact result if $J_\perp=0$
(or $J=0$).
Note, however,
that this is not a perturbation theory with respect to the interchain interaction.
The main shortcoming of the exploited approach is a poorly controlled
mean-field treatment of the phase factors
that arise after inserting (\ref{04}) into (\ref{02}).
New theoretical work is needed
to go beyond this approximation \cite{[11]}.
Another challenge is to clarify whether such a theory
is capable of capturing
the physics in spin liquid phase in Cs$_2$CuCl$_4$.
For this purpose the calculations
should be repeated for the Heisenberg model
on a triangular lattice (\ref{01})
(probably with the Dzyaloshinskii-Moriya interaction \cite{[03]}).
We have left this problem for future studies.

\end{document}